\documentstyle[12pt]{article}
\def\ifm{\ifmmode}
\def\pt{\ifm p_{\rm T} \   \else $p_{\rm T}$\ \fi}

\newcommand{\notp}{\ \hbox{{$p$}\kern-.43em\hbox{/}}}
\newcommand{\beq}{\begin{eqnarray}}
\newcommand{\eeq}{\end{eqnarray}}
\def\PL #1 #2 #3 {{\it Phys. Lett.} {\bf#1} (#3) #2}
\def\NP #1 #2 #3 {{\it Nucl. Phys.} {\bf#1} (#3) #2}
\def\PR #1 #2 #3 {{\it Phys. Rev.} {\bf#1} (#3) #2}
\begin{document}
\baselineskip=14pt
\begin{flushright}
FSU--HEP--930608\\
MAD/TH/93/4\\
MAD/PH/769\\
Board: hep-ph/9307227\\
June 1993
\end{flushright}
\begin{center}
{\bf MEASUREMENT OF THE STRANGE QUARK DISTRIBUTION FUNCTION
 IN W + CHARM QUARK EVENTS}
\footnote{Talk presented by S. Keller at the workshop``Physics at
Current Accelerators and the Supercollider", Argonne National
Laboratory, June~2 --~5, 1993, to appear in the proceedings.}
\vglue .50cm
{U.~Baur, S.~Keller\\}
{\it Department of Physics, B--159, Florida State University\\}
{\it Tallahassee, Florida 32306, USA,\\}
\vglue 0.3cm
{and\\}
\vglue 0.3cm
{F.~Halzen, K.~Riesselmann\\}
{\it Department of Physics, University of Wisconsin,\\}
{\it Madison, Wisconsin 53706, USA,\\}
\vglue 0.3cm
{and\\}
\vglue 0.3cm
{M.~L.~Mangano\\}
\baselineskip=13pt
{\it INFN, Scuola Normale Superiore and Dipartimento di Fisica,\\}
{\it Pisa, Italy\\}
\vglue 0.8cm
\end{center}
\begin{abstract}
\noindent
We investigate the prospects of measuring the strange quark
distribution function at the Tevatron, using $W$ plus charm quark
events. The $W$ plus charm quark signal produced by strange
quark--gluon fusion, $sg\rightarrow W^-c$ and $\bar sg\rightarrow W^+\bar c$,
is approximately 5\% of the inclusive $W+1$ jet cross section for jets
with a transverse momentum $p_T(j)>10$~GeV. We study the sensitivity of
the $W$ plus charm quark cross section to the parametrization of the
strange quark distribution function, and evaluate the various
background processes. Strategies to identify charm quarks in CDF and
D$0\llap/$ are briefly described.
\end{abstract}
\newpage
\section{Introduction}
One of the main goals of deep inelastic scattering experiments is to
obtain reliable measurements of the parton distribution functions of
the proton. Recently, the CTEQ\cite{CTE93} and CCFR\cite{CCF92}
collaborations have determined the ratio of momentum fractions of
strange quark versus $\bar u$ plus $\bar d$ quarks, $\kappa =2S/(\bar
U+ \bar D)$. The CCFR collaboration found $\kappa\approx 0.4$ in its
analysis of di--muon events in deep inelastic scattering, whereas the
CTEQ collaboration with its global fit analysis obtained a result of
$\kappa\approx 1$.  The difference between the two results is at the
$3-4\sigma$ level.  Among current parametrizations of the parton
distribution functions the MRSD0 set\cite{MRS93} best represents the
CCFR result. The value of $\kappa$ in this set was fixed to
$\kappa=0.5$ at $Q^2=4$~GeV$^2$. As a representative of the CTEQ
result, we shall use the CTEQ1M set in the following. The difference
between the two parametrizations in $\kappa$ is approximately a factor
of two at low $Q^2$, and decreases with increasing $Q^2$. This is
illustrated in Fig. 1, where the ratio of the strange quark
distribution functions for the CTEQ1M and MRSD0 set for $Q^2=
5$~GeV$^2$ and $Q^2=M_W^2$ is shown.

\begin{figure}[h]
\vskip 3in
\caption{Ratio of the strange quark distribution functions for the
CTEQ1M set and the MRSD0 set for two different values of $Q^2$.}
\end{figure}
\newpage
At large $Q^2$, the strange quarks that originate from gluon splitting have
comparable distribution for the two sets because the two gluon distribution
functions are very similar.

\begin{sloppypar}
Here, we point out that the discrepancy between the CTEQ and CCFR
result could be resolved by a direct, independent measurement of the
strange quark distribution function. At the Tevatron, such a
\end{sloppypar}
measurement could be carried out by determining the charm content of
$W+1$~jet events. The paper is organized as follows. In Section 2, the
signal and the leading background processes are studied.  In Section 3,
different experimental techniques to tag a charm  quark inside a jet
are discussed. Finally, in Section 4, we present our conclusions
together with some additional remarks.

\section{Signal and Background}
Associated $W+$~charm production proceeds, at lowest order, through
$sg$ and $\bar sg$ fusion, $sg\rightarrow W^-c$ and $\bar sg\rightarrow
W^+\bar c$. The alternative process where the $s$ quark in the reaction
is replaced by a $d$ quark, is suppressed by the quark mixing matrix
element $V_{cd}$. This suppression is somewhat compensated by the
larger $d$ quark distribution function, such that the $dg\rightarrow
Wc$ cross section is about 10\% of the $sg\rightarrow Wc$ rate.  Since
the final state is identical for these two subprocesses, the sum of the
$sg$ and $dg$ contributions will be considered as the ``signal'' in the
following.

The largest background originates from the production of a $c\bar c$
pair in the jet recoiling against the $W$.  If only the $c$, or the
$\bar c$ quark, is identified in the jet, the event looks like a signal
event. Similarly, a $b\bar b$ pair can be produced in the jet, and the
$b$ or the $\bar b$ quark misidentified as a charm quark.  Assuming
that all the $b$ quarks are misidentified, approximately 75\% (20\%) of
the background originates from a $c\bar c$ ($b \bar b$) pair produced
in a jet initiated by a gluon. The remaining 5\% is due to the
production of a $c \bar c$ pair in a quark--initiated jet.

For our subsequent discussion it is convenient to define the following
two ratios of cross sections:
\beq
R_1=\frac{signal}{W+1~jet},
\eeq
and
\beq
R_2 = \frac{signal+background}{W + 1~jet},
\eeq
where ``$background$'' includes the three background processes
mentioned earlier, and ``$W + 1~jet$'' refers to the total inclusive
$W$ + 1~jet cross section. $R_1$ represents the ideal situation in
which all background events are completely eliminated. $R_2$ describes
the case where none of the background is removed. In practice, the
ratio of events with a tag for a charm quark inside of the jet over the
total number of events will be measured. The result for this ratio will
fall in between $R_1$ and $R_2$, because specific methods of tagging
the charm quark inside of the jet will suppress some part of the
background, see Section 3. There are three possibilities to reduce the
background:

\begin{itemize}
\item Charge reconstruction: for the signal, the $W$ and $c$
quark electric charges are correlated. For the $c\bar c$ background, the
charm quark has the wrong charge 50\% of the time. Therefore, if the charges of
the $W$ and of the $c$ quark are determined, the $c\bar c$
background can be reduced by
a factor of two.  Furthermore, events with the wrong charge
correlation provide a measurement of the background,
that could subsequently be subtracted.

\item Cut on the charm quark transverse momentum: since more than
one charm quark is present in the background processes, the average $p_T$
of the charm quark is smaller than in the signal.
\item Flavor identification: if the bottom quark is
identified, the $b\bar b$ background can be subtracted.
\end{itemize}

To numerically simulate the signal and background processes we have used
the Monte--Carlo program PYTHIA\cite{PYT87} (version 5.6). For
$W+c\bar c$ and $W+b\bar b$ production, we have compared the result of
PYTHIA with that of the matrix element calculation of Ref.~\ref{Man92}. The
results of both calculations are in general agreement, with PYTHIA resulting
in a somewhat larger background. In our simulations, a ``jet'' is defined in
the following way. The direction of the sum of the momenta of all the partons
produced in the shower is taken as the center of a cone of radius
$\sqrt{\Delta\eta^2 +\Delta\phi^2} = 0.7$, where $\eta$ is the pseudorapidity
and $\phi$ the azimuthal angle. All the partons inside the cone are
considered to be part of the jet. The $W$ is assumed to decay into an
electron (or
positron) and a neutrino. To simulate the acceptance of a real detector, we
impose the following cuts on the final state particles:
\beq
p_{T}(e) \geq 20~{\rm GeV}, \hskip 1.cm |\eta(e)| \leq 1.  \nonumber    \\
\notp_{T} \geq 20~{\rm GeV}, \hskip 1.cm \phantom{|\eta(e)| \leq 1.}
\\
p_{T}(j) \geq 10~{\rm GeV}, \hskip 1.cm |\eta(j)| \leq 1. \nonumber
\eeq
Our results for $R_1$ and $R_2$ do not depend sensitively on the cuts.
Events with two charm quarks inside a jet are counted
twice. Numerical values for
$R_1$ and $R_2$ are presented in Table~\ref{tab:rat} for the MRSD0
and CTEQ1M parametrizations.
\begin{table}[h]
\centering
\caption
[Ratios.]
{The ratios $R_1$ and $R_2$ for the MRSD0 and CTEQ1M parametrization.
}
\vspace{2.mm}
\begin{tabular}{|c|c|c|}\hline
&$R_1(\%)$& $R_2(\%)$  \\ \hline\hline
MRSD0  &4.2    &8.4    \\ \hline
CTEQ1M &5.3    &9.7    \\ \hline
\end{tabular}
\label{tab:rat}
\end{table}

As can be seen in Table~\ref{tab:rat}, $R_2$ is about twice as large as $R_1$,
which means that the signal-to-background ratio is of order unity.
The average values for the ratios $R_1$ and $R_2$ are roughly 5\% and 9\%,
respectively.  To quantify the difference between
the two sets, we define the ratio
\beq
\Delta=\frac{R(CTEQ1M)-R(MRSD0)}{\frac{1}{2} (R(CTEQ1M)+R(MRSD0))},
\eeq
where $R$ stands for $R_1$ or $R_2$. $\Delta$ is equal to $23\%$ and $14\%$
for $R_1$ and $R_2$, respectively.
In order to perform a meaningful measurement,
the experimental uncertainty in measuring $R_1$ or $R_2$ must
be less than the corresponding value for $\Delta$.
In Fig. 2, the ratios $R_1$ and $R_2$ are
shown for the two parametrizations as a function of the \pt of the
jet.
\begin{figure}[t]
\vspace{3in}
\caption{The cross section ratios $R_1$ and $R_2$ as a function of the
jet $p_T$ for the MRSD0 and CTEQ1M parametrizations of the parton
distribution functions.}
\end{figure}

The two sets of parton distribution functions yield the same values for
the inclusive $W+1$~jet cross section and
the background to within 1\%. The variation of the ratios $R_1$ and $R_2$
with the parton distribution functions therefore directly reflects the
difference in the strange quark distribution. Not surprisingly, both cross
section ratios are not very sensitive to changes in the factorization
scale $Q^2$.
Varying $Q^2$ between 1/4 and 4 times the default average $Q^2$ of PYTHIA, the
ratios change only by $\Delta R/R\approx4$\%. The
stability of $R_1$
and $R_2$ with respect to variations in $Q^2$ indicates that the sensitivity
of the ratios to the strange quark distribution function is unlikely to be
overwhelmed by uncertainties originating from higher order QCD corrections.

Assuming both electron and muon decay channels of the $W^{\pm}$,
an integrated luminosity of 10~pb$^{-1}$ yields about 2000
$W+1$~jet events for the cuts summarized in Eq.~(3). This corresponds to
approximately 100 $W$ plus charm quark signal events, and to about the
same number of potential background events. From the expected number of
signal events it is straightforward to estimate the charm tagging
efficiency, $\epsilon_c^{min}$, required to be statistically sensitive to the
variation of the $W + c$ production cross section with the strange quark
distribution function. Depending on how efficiently the various background
processes can be suppressed, we find $\epsilon_c^{min}\approx 20 - 30$\%
for an integrated luminosity of 10~pb$^{-1}$. Note that
$\epsilon_c^{min}$ scales like $(\int\!{\cal L}dt)^{-1}$.

\section{Charm Quark Tagging in CDF and D$0\llap/$}

CDF and D$0\llap/$ explore three different strategies to identify charm
quarks:

\begin{enumerate}

\item Search for a displaced secondary vertex in the vertex
detector. The efficiency to tag $b$ quarks with the SVX\cite{SVX90} of
CDF is about 10--20\%, depending on the $p_T$ range.
The tagging efficiency for charm quarks is
expected to be smaller than that for bottom quarks, due to the smaller
mass and decay track multiplicity of the charmed hadrons.

\item Reconstruction of exclusive nonleptonic charmed baryon or
meson decays.
CDF, for example, uses the decay channel $D^0 \rightarrow K \pi$ to
identify semileptonic B meson decays\cite{CDF93}.
Other exclusive channels will be added in the future, and an efficiency of a
few percent should be reached.
\item Searching for inclusive semileptonic charm decays~\cite{LEP93}.
The average inclusive semileptonic charm decay branching ratio is
$B(c\rightarrow e\nu,\mu\nu)\sim 10$\%. If one assumes
a reconstruction efficiency for a
muon inside a jet of the order of 50\%, a total charm tagging efficiency
from semileptonic charm decays of the order of 5\% may well be possible.
\end{enumerate}

Combined, the three methods may yield an overall charm detection
efficiency of about 10\%.

\section{Conclusions}

We have studied the prospects for measuring the
strange quark distribution function in $W+c$ production at the Tevatron.
The method is similar to the one described in Ref.~\ref{Fle89}
for measuring the charm quark distribution function in $\gamma$ plus
charm  production.
Our results indicate that, for the data sample accumulated in the
1992--93 run, the expected charm tagging efficiencies are a limiting
factor. For 100~pb$^{-1}$, however, one could seriously attempt to
determine the strange quark distribution function from $W+c$ production.

In our analysis we have concentrated on the charm content of $W+1$~jet
events with $p_T(j) > 10$~GeV. Alternatively one could search for $W+c$
production in the inclusive $W$ sample, without requiring the presence
of a high transverse momentum jet. The advantage here would be a
significant increase of the number of signal events.
However,
due to the smaller average transverse momentum of the charm quarks
in the inclusive $W$ sample, the charm quark detection efficiency is
expected to be reduced.

Clearly, more experimental and theoretical work is needed in
order to determine reliably whether a measurement of the $s$ quark
distribution function in $W+c$ production is feasible.

If the strange quark distribution function is measured precisely in
other experiments, $W$ plus charm quark production may eventually be used
to measure the quark mixing matrix element $V_{cs}$ at high $Q^2$ and
compare it with the value extracted from low energy
experiments\cite{Grz87}.

\section{Acknowledgments}

We are grateful to J.~Bla\-zey,
S.~Erre\-de, B.~Flaugher, B.~Fletcher, T.~Heuring, T.~Lecompte,  H.~Reno, and
M.~L.~Stong for useful and stimulating discussions.  This research was
supported in part by the  Texas National
Research Laboratory Commission under grant no. RGFY9273, and by the
U.S. Department of Energy under
contract numbers DE-FG05-87ER40319, and DE-AC02-76ER00881.

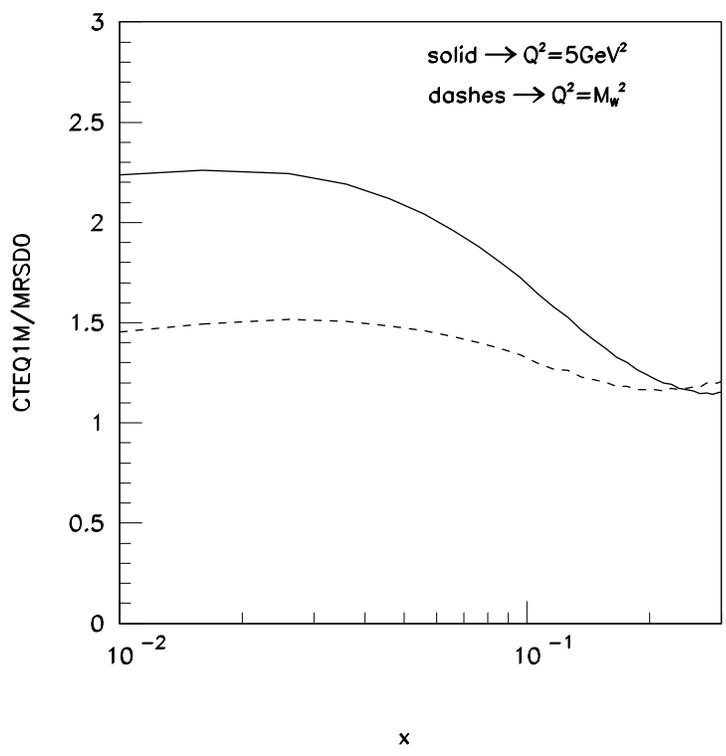
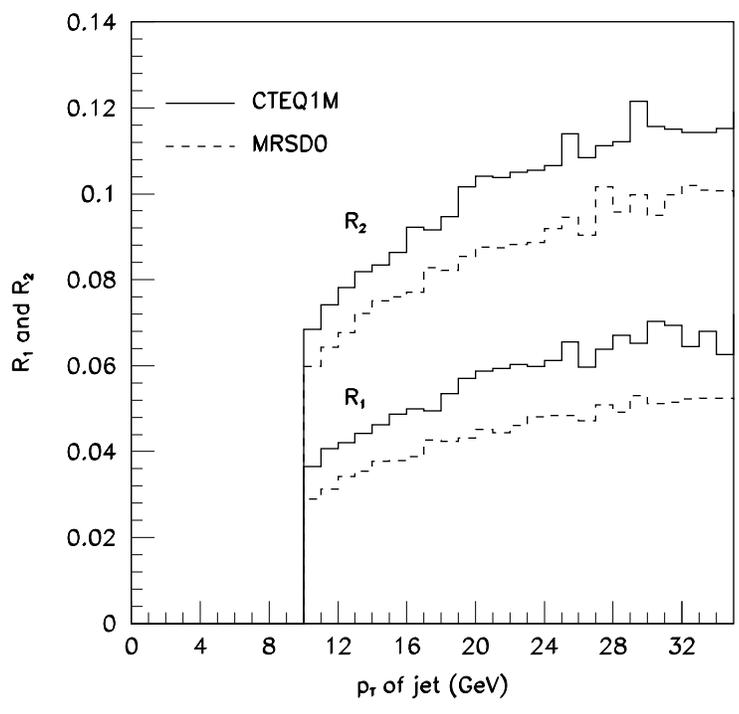
\end{document}